%% file: main.tex
\documentclass[10pt,conference]{IEEEtran}
\IEEEoverridecommandlockouts
\usepackage[T1]{fontenc}
\usepackage{amsmath,amssymb}
\usepackage{graphicx}
\usepackage{booktabs}
\usepackage{multirow}
\usepackage{array}
\usepackage{enumitem}
\usepackage{csquotes}
\usepackage{tcolorbox}
\usepackage[numbers]{natbib}
\usepackage{xcolor}
\usepackage{colortbl}
\usepackage{tikz}
\usetikzlibrary{calc}
\usepackage{pgfplots}
\pgfplotsset{compat=1.18}
\usepackage{siunitx}
\usepackage{etoolbox}
\usepackage{makecell}
\usepackage{tabularx}
\usepackage{longtable}
\usepackage{framed}
\usepackage{soul}
\usepackage{float}
\usepackage{caption}
\usepackage{subcaption}
\usepackage{refcount}
\usepackage{url}
\usepackage[hidelinks]{hyperref}
\usepackage{microtype}
\def\BibTeX{{\rm B\kern-.05em{\sc i\kern-.025em b}\kern-.08em T\kern-.1667em\lower.7ex\hbox{E}\kern-.125emX}}

\definecolor{quotepurple}{RGB}{94,52,140}
\definecolor{TekheletPurple}{HTML}{572F96}
\definecolor{vlightgray}{gray}{0.9}
\newcommand{\pq}[1]{{\color{quotepurple}\enquote{#1}}}

\tcbset{takeaway/.style={colback=black!4,colframe=black!55,boxrule=0.4pt,arc=1.5pt,left=4pt,right=4pt,top=3pt,bottom=3pt,before skip=6pt,after skip=6pt}}
\newif\ifdraft
\draftfalse 
\newif\ifrevising
\revisingfalse

\newcommand{\deleted}[1]{{\ifrevising{\relax}\else\relax\fi}}

\newcolumntype{P}[1]{>{\raggedright\arraybackslash}p{#1}}

\makeatletter
\providecommand\nopunct{\@addpunct{}}
\makeatother

\definecolor{gInkDV}{HTML}{35577D}\definecolor{gInkDP}{HTML}{8A6A22}\definecolor{gInkQR}{HTML}{9C3D30}
\definecolor{gInkIO}{HTML}{0E7C9E}\definecolor{gInkMW}{HTML}{C2661B}\definecolor{gInkCC}{HTML}{3B3F46}
\newcommand{\gl}[2]{\textcolor{gInk#1}{\textbf{#1#2}}}

\usepackage[normalem]{ulem}

\begin{document}
\bstctlcite{IEEEexample:BSTcontrol} 

\title{``What Should I Do?'' Guidelines for Responsibly Integrating AI in Software Engineering\vspace{-0.3\baselineskip}}

\author{\IEEEauthorblockN{Rudrajit Choudhuri\IEEEauthorrefmark{1},
Fatima A. Moussaoui\IEEEauthorrefmark{1},
Carmen Badea\IEEEauthorrefmark{2},
Christian Bird\IEEEauthorrefmark{2},
Jenna L. Butler\IEEEauthorrefmark{2},
Brian Houck\IEEEauthorrefmark{3},\\
Margaret Burnett\IEEEauthorrefmark{1},
Anita Sarma\IEEEauthorrefmark{1}}
\IEEEauthorblockA{\IEEEauthorrefmark{1}Oregon State University, Corvallis, OR, USA\\\{choudhru, moussaof, anita.sarma\}@oregonstate.edu, burnett@eecs.oregonstate.edu}
\IEEEauthorblockA{\IEEEauthorrefmark{2}Microsoft, Redmond, WA, USA, \{cabadea, cbird, jennbu\}@microsoft.com\quad \IEEEauthorrefmark{3}DX, brian.houck@getdx.com}}

\IEEEaftertitletext{\vspace{-1.1\baselineskip}} 
\maketitle

\begin{abstract}

Speculations abound about the ``right'' ways for developers to integrate AI into their software engineering (SE) work, but the evidence that exists is sparse and contradictory.
Yet, developers are held responsible for the products they ship with (or without) AI's help, so they need to integrate AI in responsible ways.
To address this gap, we conducted a two-phase, large-scale investigation with 1,118 developers at Microsoft, GitHub, and Pfizer.
In Phase~1, we surveyed 860 developers at Microsoft on where they sought and resisted AI support in their daily work, and from their accounts, synthesized 35 developer-facing guidelines: 29 across five categories of SE work and 6 cross-cutting ones.
A year later, as AI capabilities and developers' experiences co-evolved, we validated them in Phase~2 with 258 developers in technology (Microsoft, GitHub) and biopharmaceutical (Pfizer) domains.
Developers in our study found the guidelines clearly worded (92\%) and relevant to their work (91\%), and most wanted to adopt them in their own work (89\%).
We also report how developers described operationalizing these guidelines in practice: using AI to refine their craft and delegate verifiable work, while keeping judgment and accountability for themselves.
For relationship-centric work, endorsement varied with developers' domain and their SE and AI experience.
Overall, we offer concrete advice from in-the-trenches developers to other in-the-trenches developers, and a shared vocabulary that engineering teams can adopt to integrate AI well in their work.
\end{abstract}

\input{sec/1.intro}
\input{sec/2.Background}

\input{sec/3.Method-new}
\input{sec/3b.Threats}

\input{table/tab_guidelines_scenarios} 
\input{sec/4.Results-new}

\input{sec/5.Discussion}

\input{sec/8.conclusion}

\bibliographystyle{IEEEtranN}
\footnotesize{\bibliography{acmart}}

\end{document}

%% file: sec/1.intro.tex
\section{Introduction}
\label{sec:intro}

Software developers have never had more help, and less certainty about how to use it, at the same time.
AI tools have gone from completing a line of code to writing whole features, reviewing pull requests, and drafting messages to customers, and most developers now use them routinely at work~\cite{dora2025}.

Each new release brings the same question back in a new form: \textit{how much of my daily work should I hand to AI, and what must I still own?}
Most developers are left to decide this on their own, and with uneven results---even within one team, some developers have worked out what to hand off and where AI cannot help, while others struggle to see where it applies at all~\cite{miller2025maybe}.
Yet whatever AI contributes, developers must remain accountable for the result, so how they answer that question shapes the quality of what they ship and their standing with the people who rely on their work.

Moreover, AI strains practices that teams built over decades, practices that assumed a person did the work and served as signals of good work.
Code review, for example, assumes that authors understand what they contributed; with AI-written code, reviewers face more contributions than they can absorb, from authors who may not fully understand them~\cite{baltes2026endless,storey2026from}.
Documentation assumes that someone vetted what it says; agents now write much of it, often merged with little human review~\cite{yamasaki2026who}.
As these signals weaken, the effort AI appears to save in code generation shifts downstream~\cite{choudhuri202622systems}, which may strain developers' well-being~\cite{guizani2026cost}, without delivering the expected productivity gains~\cite{becker2025measuring}.

When AI entered consumer products, the field responded with guidelines for how AI systems should be designed to interact with their users~\cite{amershi2019guidelines,weisz2024design}.
Developers now stand on the other side of that relationship, deciding how to integrate AI responsibly into their own work.
Speculation about the ``right'' ways abounds, but no shared guidance, drawn from developers in the trenches for other developers in the trenches, says where AI should help in software engineering (SE), where it should stay out, and what developers should demand of such tools.

In this paper, we synthesize what \textbf{860} developers at Microsoft, a frontier technology company whose developers both build and use AI, had worked out into guidelines for responsibly integrating AI into software work.
In 2025, we asked these developers to describe where they sought and resisted AI across SE tasks (see Table~\ref{tab:SDLC-categories}), and from their accounts, we synthesized 35 guidelines: 29 across different categories of SE work, alongside 6 cross-cutting ones that apply across categories.
A year later, as AI capabilities and developers' experiences co-evolved, we validated the guidelines beyond the domain they came from, with \textbf{258} developers at Microsoft, its subsidiary GitHub, and Pfizer, a company in a regulated biopharmaceutical domain, who rated each guideline on its clarity, its relevance to their own work, and whether they wanted to adopt it, and described how they would operationalize it in practice.
We waited a year on purpose, a considerable period given how rapidly AI tools are advancing, to test whether the guidelines remained durable as the mechanics of developers' day-to-day work changed, on the premise that what developers fundamentally do at work (e.g., software development, design, infra \& ops, and quality assurance) endures far longer.

Overall, the main contribution of this paper is a set of developer-facing guidelines for responsibly integrating AI into daily SE work, validated across companies and domains, and reported alongside the practices developers described for operationalizing each one.
The guidelines can give engineering teams a shared vocabulary for deciding where AI belongs in their context and how to use it with more certainty, all while remaining responsible for what they ship.

%% file: sec/2.Background.tex
\section{Related Work}
\label{sec:backg}

\label{sec:related-AImatters}
As AI tools enter development workflows, SE research has examined how developers use them~\cite{liang2024large,bird2022taking}, what drives their adoption~\cite{russo2024navigating,lambiase2025exploring,choudhuri2025guides,choudhuri2025needs}, and what they cost~\cite{guizani2026cost,becker2025measuring}.
Literature reviews catalog a fast-growing range of SE tasks that AI tools now support~\cite{hou2024large,fan2023large}, and early industry accounts described the opportunities of AI pair programmers~\cite{bird2022taking}.
Large-scale surveys followed, documenting successes and recurring usability challenges of AI tools~\cite{liang2024large}, while controlled and field studies measured their effects on productivity~\cite{peng2023impact,ziegler2024measuring,butler2025dear}.

Capability alone, however, does not explain adoption.
\citet{russo2024navigating} found that compatibility with developers' existing workflows drove early AI adoption, and \citet{lambiase2025exploring} showed that individual factors shape adoption differently across SE tasks.
Trust, shaped by both the tool and the developer, further governs whether developers rely on AI's output~\cite{johnson2023make,wang2023investigating,cheng2023would,choudhuri2025guides,choudhuri2025needs}.
Individual and team factors also drive how much developers use these tools~\cite{miller2025maybe}, just as peer influence has been shown to spread the adoption of other developer tools~\cite{xiao2014social,witschey2015quantifying}.

More recently, task-level studies found that developers tend to want AI for repetitive work such as testing and documentation but are less receptive to it in collaborative and creative work~\cite{khemka2024toward,kumar2025time,lambiase2025exploring,pereira2025exploring}.
In our prior work, developers sought AI support to reduce toil and amplify their craft, and limited it in identity-defining and relationship-centric work~\cite{choudhuri2026aiWhereMatters}.
Our subsequent investigation identified where and why developers draw the line on AI autonomy~\cite{choudhuri2026you}, the AI systems they want built next, and the constraints that make such systems acceptable in practice~\cite{choudhuri202622systems}.

Other research has investigated the costs of integrating AI poorly.
When AI output drifts from developers' goals, it costs them effort to prompt, verify, and modify it, with consequences for their trust and adoption~\cite{choudhuri2025needs,liang2024large}.
Some AI-generated work that appears to save costs simply pushes them downstream: generating code faster moves the burden onto review, documentation, and maintenance~\cite{choudhuri202622systems,baltes2026endless,storey2026from}, which may eventually strain developers' well-being~\cite{guizani2026cost} and erode the promised productivity gains~\cite{becker2025measuring}.

%
%
\label{sec:related-guidelines}
To help address problems arising from the influx of AI into software, guidelines have begun to emerge.
For example, to address problems in user-facing aspects of AI, \citet{amershi2019guidelines} synthesized 18 guidelines for human--AI interaction from more than 150 design recommendations and validated them with design practitioners.
\citet{weisz2024design} articulated user-experience (UX) design principles for generative AI applications, and \citet{liao2022designing} examined how systems should communicate their trustworthiness.

Guidelines for ethically responsible AI have also emerged in industry guidebooks and principles~\cite{GoogleGuidelines, GoogleAIPrinciples2023, MicrosoftRAI2025} and in patterns that operationalize it for teams that engineer AI systems~\cite{lu2024responsible}.
Similar guidelines have also emerged for empirical SE research involving AI-integrated systems~\cite{wagner2025towards}.
Beyond individual products and practices, the European Commission's High-Level Expert Group on AI issued ethics guidelines for trustworthy AI, setting out seven requirements, such as human agency and oversight, transparency, and accountability, for those who develop and deploy AI systems~\cite{european2019ethics}.

Still, none of this research tells software developers how to integrate AI responsibly into their own work.
Our work fills this gap with empirically grounded guidelines, synthesized from the accounts of hundreds of developers about their daily work and validated across companies and domains.

%% file: sec/3.Method-new.tex
\section{Method} 
\label{sec:method}

Our goal was to derive guidelines that help developers responsibly integrate AI into their daily work.
To do so, we conducted a two-phase investigation following the approach of \citeauthor{amershi2019guidelines}'s guidelines for human--AI interaction~\cite{amershi2019guidelines}.
In Phase~1, we synthesized our guidelines from 860 developers at Microsoft on where, why, and how they wanted AI support in their work (Sec.~\ref{subsec:phase1}).
In Phase~2, we validated them with an independent sample of 258 developers at Microsoft, GitHub, and Pfizer, spanning technology and biopharmaceutical domains (Sec.~\ref{subsec:phase2}).
Because integration is learned in practice, we used developers' accounts of their own work to ground these guidelines empirically.

\vspace{-1mm}
\subsection{Phase 1: Synthesizing the Guidelines}
\label{subsec:phase1}

\subsubsection{\textbf{Source survey}}
\label{subsubsec:phase1-survey}
We collected the Phase~1 data in our 2025 \textit{AI Where It Matters} survey, designed to characterize developers' experiences and their openness to and use of AI, reported in~\cite{choudhuri2026aiWhereMatters}, and to synthesize guidelines for responsibly integrating AI into their work, which is the focus of this paper.

The survey invited 8,000 developers across product groups, roles, and geographies, and ultimately received 1,193 responses (a 14.9\% response rate, consistent with prior SE surveys~\cite{punter2003conducting,storey2019towards}), of which 860 were valid (details in~\cite{choudhuri2026aiWhereMatters}).
In summary, the survey spanned 20 SE tasks in five categories from a grounded taxonomy of developers' work (Table~\ref{tab:SDLC-categories}).
Participants selected up to three categories that best reflected their work.
To reduce fatigue, the meta-work category, which applies to all developers, was shown only to those who selected two.
In each category, participants answered Likert-scale items on their experiences across its tasks and their openness to and use of AI support~\cite{choudhuri2026aiWhereMatters}, followed by two open-ended questions: \textit{Thinking about how to integrate AI into your work responsibly...}
\begin{enumerate}[nosep]
    \item \textit{Integration:} ``\textit{...where do you want AI to play the biggest role in [category] activities?}''
    \item \textit{Resistance:} ``\textit{...what aspects do you not want AI to handle in [category] activities and why?}''
\end{enumerate}


\input{table/SDLC-categories}

The 860 participants provided 2,580 responses to these questions: Development (816), Design \& Planning (548), Meta-work (532), Quality \& Risk (401), and Infrastructure \& Operations (283).
Note that we intentionally left \textit{responsibly integrating AI} broad, leaving it to the individual respondents to interpret it according to their own professional values.

\subsubsection{\textbf{Data analysis}}
To derive the guidelines, we analyzed participants' open-ended responses using reflexive thematic analysis~\cite{braun2006using, braun2022conceptual}.
Four authors took part in this process, comparing and contrasting codes over multiple meetings and resolving differences through negotiated agreement~\cite{creswell2016qualitative, mcdonald2019reliability}.
The unit of analysis was one participant’s responses to both questions in one category, and these responses could receive multiple codes.
In the \textit{AI Where It Matters} study~\cite{choudhuri2026aiWhereMatters}, we had already mapped each response to the tasks it addressed.
During this current paper's analysis, two authors inductively coded the responses to both questions for each task, this time to derive guidelines, and the team refined the codes into themes of where developers wanted to integrate or resist AI in their work, yielding 105 task-level guidelines (typed as \textsc{Integration}/\textsc{Resistance}).
Finally, we grouped task-level guidelines that shared the same essence into 29 guidelines across the five task categories, and those that recurred across categories into 6 cross-cutting guidelines.
The supplement links each guideline to its task-level guidelines~\cite{supplemental}.

These guidelines take three forms.
\textit{\textbf{Use-and-demand}} guidelines pair a use of AI with what developers said they demand of it in that use (``\textit{Use AI to \ldots; demand AI to \ldots}'').
\textit{\textbf{Limit-or-prohibit}} guidelines name where developers said AI's role in that category should be limited or prohibited (``\textit{Limit AI's role in \ldots}'', ``\textit{Prohibit AI from \ldots}'').
\textit{\textbf{Cross-cutting}} guidelines contain developers' demands that apply across all categories, alongside the category-level guidelines (``\textit{Demand AI to \ldots}'').

\textbf{Expert review.}
Before Phase~2 validation, we sent the guidelines to 12 experienced software developers and 5 empirical SE researchers outside the research team, with an average of 10 years of industry experience.
From their feedback, we made minor wording revisions and also included example scenarios for each guideline, grounded in its task-level origins, and offered them on demand through info-icons as part of the Phase~2 survey (see supplement~\cite{supplemental}).

\vspace{-1mm}
\subsection{Phase 2: Validating the Guidelines}
\label{subsec:phase2}

\subsubsection{\textbf{Survey design}}
To validate the guidelines synthesized in Phase~1, we surveyed an independent sample of developers within and outside the technology domain.
The Phase~2 survey followed the same structure as Phase~1, with two deliberate choices in fielding it.
First, we waited a year to test whether the guidelines remained durable as AI tools and developers' practices evolved.
Second, we deployed it in more than one domain, at Microsoft (\& GitHub) and at Pfizer, to test whether the guidelines applied beyond the technology domain.

Participants first reported their SE experience, AI experience, and, optionally, gender.
They then selected the kinds of work they had focused on in the last three months, using the Phase~1 categories, and rated each guideline in up to three selected categories, plus the cross-cutting ones, on three statements:
\begin{itemize}[nosep]
    \item \textbf{C}larity: ``\textit{This is clearly worded}'';
    \item \textbf{R}elevance: ``\textit{This applies to my work}'';
    \item \textbf{A}doption: ``\textit{I want to adopt this in my work}''.
\end{itemize}
To mitigate ordering effects, we randomized the order of the guidelines within each block.
We instructed participants to rate each guideline such that agreement meant they agreed with every part of it.
Relevance and adoption also offered an Unsure option so undecided participants were not forced to choose; we report Unsure responses (6\% of all ratings) in the supplement~\cite{supplemental}.
We did not offer the Unsure option for clarity, since participants could directly judge whether a guideline was clearly worded.
Each block also included two optional open-ended questions, which we used to analyze how participants operationalized the guidelines in their own work: an application question (``\textit{How would you actually use one of these in [category] work?}'') and a reflection question (``\textit{Is there anything else you'd like to share about the ways of using AI in [category] work you rated above?}'').

We piloted the survey with 25 developers at Microsoft.
Based on their feedback, we replaced an originally planned five-point scale with Yes/No/Unsure options to reduce cognitive burden, which also gave a clear decision on each guideline.
Both surveys appear in the supplement~\cite{supplemental}.

\subsubsection{\textbf{Participants}}
We administered our survey in Qualtrics. 
At Microsoft, we invited 4,000 developers over email, sampled uniformly at random across product groups, experience, domains, and geographies, excluding Phase~1 participants.
We sent one reminder after a week and received 332 responses (an 8.3\% response rate).
To comply with organizational requirements, we sent a single survey link that reached both Microsoft and its subsidiary GitHub, so we cannot separate their participants and hereafter refer to both as Microsoft.
Beyond Microsoft, we reached out to five other companies across different domains, and Pfizer agreed to participate.
At Pfizer, we worked with an engineering leader to distribute the survey internally, who then cascaded it to team members; for this reason, we cannot report a response rate.
Participation was voluntary and anonymous in all cases, and the survey remained open from August to September 2026; the institutional review boards at Microsoft and Oregon State University approved our study.\looseness=-1

Of the 406 responses (332 at Microsoft and 74 at Pfizer), we excluded those that rated no guidelines (MS: 93; PF: 30) or had implausibly fast completion times (MS: 22; PF: 3).
We retained 258 participants, 217 at Microsoft (MS1--MS217) and 41 at Pfizer (PF1--PF41).
Participants spanned a wide range of SE and AI experience.
Most self-identified as men (85\%) and rated themselves as highly experienced with AI tools (60\%).
Participant demographics are provided in the supplement~\cite{supplemental}.

\subsubsection{\textbf{Analysis of ratings}}
\label{subsubsec:ratings}
For each guideline and statement, we report the percentage of agreement (Yes among Yes and No responses), with 95\% Wilson score intervals in the supplement; category-level and overall percentages pool the ratings in each category and across all guidelines.
We retained all guidelines, as a majority of participants found each clear and relevant and wanted to adopt it.
We mark 70\% on each bar in Table~\ref{tab:guidelines-scenarios}, a commonly accepted level of consensus~\cite{stemler2004comparison}, and star ($^\star$) the guidelines whose adoption fell below it.

Domain comparisons (Microsoft vs.\ Pfizer; Benjamini--Hochberg-corrected Fisher's exact tests; see~\cite{supplemental}) showed no statistically significant difference. We thus report domain differences descriptively where relevant, in percentage points.

\subsubsection{\textbf{Analysis of open-ended responses}}
\label{subsubsec:openended}
We analyzed 726 responses from the 168 participants who answered the open-ended questions to understand how they would operationalize each guideline, to explain their ratings, and to identify misreadings.
Because responses addressed specific guidelines, each guideline needed its own codebook.
To code consistently across them, we used a human-in-the-loop, multi-model AI council procedure following our prior work~\cite{choudhuri202622systems}, guided by reflexive thematic analysis~\cite{braun2006using,braun2019reflecting}.
The models performed bounded, auditable operations, and we retained interpretive authority~\cite{tai2024examination}.
Specifically, we decided what each theme meant and which interpretations entered the analysis, and every theme and coding decision remained traceable to participants' verbatim responses.

To build the codebooks, we first had three models from distinct provider families, Claude Opus~5.5 (Anthropic), GPT-5.6 (OpenAI), and Gemini Flash~5.7 (Google), independently discover candidate themes for each guideline operationalization, with the participant IDs and verbatim excerpts that supported them; using distinct model families reduces the risk that one model family's biases or failure modes dominate~\cite{wu2026council}.
Each model analyzed the responses through different lenses: \textit{semantic} (what participants said), \textit{latent} (the division of labor between the developer and AI that a response implied), and \textit{contrastive} (the conditions under which participants would (not) use AI as a guideline describes).

We then used GPT-5.6 to reconcile the candidates into one codebook per guideline, with a definition and inclusion and exclusion criteria for each code; one model sufficed because three had independently surfaced them.

Two authors reviewed every candidate theme against all its cited responses, assessing its \textit{fidelity} to what participants described, its \textit{grounding} in those responses, and its \textit{distinctness} for systematic coding.
Through negotiated agreement, we dropped weakly substantiated themes, merged similar ones, and split those that combined different operationalizations.

After finalizing the codebooks, we used three new model instances (Claude Opus~5.5, GPT-5.6, and Gemini Flash~5.7) to code every response independently.
Each model gave a rationale before assigning a code~\cite{wei2022chain}, leaving a traceable link between each response and its codes for human review.
We disallowed the models from introducing new codes at this point to prevent code drift.
Agreement among the three model coders was high (mean Krippendorff's $\alpha = 0.92$; per-category medians 0.89--1.00~\cite{krippendorff2018content}), which indicates coding consistency.
A response received a code when at least two of the three model coders assigned it.
As a final validation, two authors independently coded a representative sample of 70 responses ($\sim$10\%); human-council agreement was Cohen's $\kappa = 0.81$.

%% file: table/SDLC-categories.tex
\begin{table}[h]
\caption{Grounded taxonomy of SE tasks~\cite{choudhuri2026aiWhereMatters}}
\label{tab:SDLC-categories}
\fontsize{7.3}{8.5}\selectfont
\setlength{\tabcolsep}{2.5pt}
\renewcommand{\arraystretch}{1.05}
\begin{tabular}{@{}>{\raggedright\arraybackslash}p{1.9cm}>{\raggedright\arraybackslash}p{\dimexpr\columnwidth-1.9cm-2\tabcolsep\relax}@{}}
\toprule
\textbf{Category} & \textbf{Tasks}\\
\midrule
\rowcolor{gray!20} Development (DV) & Coding/Programming; Bug Fixing/Debugging; Performance Optimization; Refactoring \& Maintenance/Updates; AI~Feature~Development\\
Design \& Planning (DP) & System Design; Requirements Engineering; Project Planning \& Management\\
\rowcolor{gray!20} Quality \& Risk (QR) & Testing \& Quality Assurance; Code Review/Pull Requests; Security \& Compliance\\
Infrastructure \& Ops (IO) & DevOps (CI/CD); Environment Setup \& Maintenance; Infrastructure Monitoring; Customer Support\\
\rowcolor{gray!20} Meta-work (MW) & Documentation; Stakeholder Communication; Mentoring \& Onboarding; Learning; Research~\&~Brainstorming\\
\bottomrule
\end{tabular}
\end{table}

%% file: sec/3b.Threats.tex
\vspace{-1mm}
\subsection{Threats to Validity} 
\label{sec:threats}

As with any empirical investigation, ours had threats to validity. Our sample may not be representative, in part because some organizations did not want to publicly discuss their practices. Of the five companies beyond Microsoft and GitHub we approached, only Pfizer agreed to participate. Self-selection bias is also a possible threat, in that developers in our study may hold stronger views about AI and thus might have been more motivated to participate. Also, developers may have endorsed guidelines they would not sustain using in their daily work. 

Accuracy of our data also depends on our qualitative analysis accurately representing participants' experiences. The guidelines are our synthesis of participants' accounts, but other researchers might group the same accounts differently. Our mitigations were grounding every task-level guideline in participants' responses, resolving disagreements through negotiated agreement, having experts review the guidelines, and validating them with an independent sample of developers across companies and domains. Our use of AI to scale Phase~2's analysis may also have introduced inaccuracies. 
To mitigate this threat, we reserved all interpretive judgments for ourselves: we had three model families propose themes independently, each traced to participant IDs and verbatim excerpts; two authors then reviewed every theme against its cited responses, finalized the codebooks, reviewed the rationale models gave for their code assignments, and checked the council's coding against their own coded sample.
We also present verbatim quotes throughout the paper to let readers judge for themselves whether participants said what the themes capture.

Finally, our investigation included data from only three companies (Microsoft, its subsidiary GitHub, and Pfizer), so we do not claim to represent all developers. Instead, we offer our guidelines as a validated starting point for engineering teams to decide what is relevant to their context. The guidelines' relevance and adoption may differ in smaller organizations, open-source settings, or more strongly regulated domains. To support replication in other contexts, we have made our survey instruments available~\cite{supplemental}.

%% file: table/tab_guidelines_scenarios.tex
\definecolor{gBar}{HTML}{2F6DB5}
\definecolor{gTrack}{HTML}{E8E8ED}
\definecolor{gInk}{HTML}{1D1D1F}
\definecolor{gSoft}{HTML}{86868B}
\newif\ifgTint \gTinttrue
\definecolor{gCatDEV}{HTML}{F3F6FA}\definecolor{gCatDNP}{HTML}{F9F6EC}\definecolor{gCatQRM}{HTML}{FAF0EE}
\definecolor{gCatINF}{HTML}{E7F3F7}\definecolor{gCatMW}{HTML}{FCEFE4}\definecolor{gCatXC}{HTML}{EFF0F2}
\definecolor{gInkDEV}{HTML}{35577D}\definecolor{gInkDNP}{HTML}{8A6A22}\definecolor{gInkQRM}{HTML}{9C3D30}
\definecolor{gInkINF}{HTML}{0E7C9E}\definecolor{gInkMW}{HTML}{C2661B}\definecolor{gInkXC}{HTML}{3B3F46}
\definecolor{gKwDEV}{HTML}{35577D}\definecolor{gKwDNP}{HTML}{6E541B}\definecolor{gKwQRM}{HTML}{9C3D30}
\definecolor{gKwINF}{HTML}{0B637E}\definecolor{gKwMW}{HTML}{914C14}\definecolor{gKwXC}{HTML}{3B3F46}
\newcommand{\gtint}[1]{\ifgTint\cellcolor{gCat#1}\fi}
\newcommand{\gtintrow}[1]{\ifgTint\rowcolor{gCat#1}\fi}

\newlength{\gCatW}\setlength{\gCatW}{12pt}
\newlength{\gGuideW}\setlength{\gGuideW}{0.355\textwidth}
\newlength{\gBarW}\setlength{\gBarW}{54pt}
\newlength{\gBarLen}\setlength{\gBarLen}{32pt}
\newlength{\gExW}
\setlength{\gExW}{\dimexpr\textwidth-\gCatW-\gGuideW-\gBarW-4\tabcolsep-5pt\relax}
\newcolumntype{Q}{@{}>{\gGuideFont\raggedright\arraybackslash}p{\gGuideW}
  >{\centering\arraybackslash}p{\gBarW}
  >{\gExFont\raggedright\arraybackslash\setlength{\parfillskip}{0pt plus .7\linewidth}}p{\gExW}@{}}
\newcommand{\gGuideFont}{\fontsize{7.3}{8.0}\selectfont}
\newcommand{\gExFont}{\fontsize{6.6}{7.2}\selectfont}

\def\gBarH{2.8}\def\gBarStep{5.4}\def\gLine{70}
\newcommand{\crabar}[3]{%
  \begin{tikzpicture}[x=\gBarLen/100, y=1pt, baseline=-6.2pt]
    \foreach \v/\l [count=\i] in {#1/C,#2/R,#3/A}{%
      \pgfmathsetmacro{\yy}{-(\i-1)*\gBarStep}%
      \ifnum\v<\gLine \def\gFill{gBar!38}\def\gTxt{gSoft}\else\def\gFill{gBar}\def\gTxt{gInk}\fi
      \fill[gTrack, rounded corners=1.4pt] (0,\yy) rectangle (100,\yy-\gBarH);
      \fill[\gFill, rounded corners=1.4pt] (0,\yy) rectangle (\v,\yy-\gBarH);
      \node[anchor=east, font=\sffamily\fontsize{5}{5}\selectfont, text=gSoft, inner sep=1.2pt] at (0,\yy-\gBarH/2) {\l};
      \node[anchor=west, font=\sffamily\fontsize{6}{6}\selectfont, text=\gTxt, inner sep=1.4pt] at (100,\yy-\gBarH/2) {\v};
    }
    \draw[gInk!80, line width=.5pt] (\gLine,1.6) -- (\gLine,{-2*\gBarStep-\gBarH-1.6});
  \end{tikzpicture}}

\newcommand{\grow}[6]{\textbf{#1}~#2 & \crabar{#3}{#4}{#5} & #6 \\}
\newcommand{\gsep}{\arrayrulecolor{black!25}\specialrule{.3pt}{.2pt}{.2pt}\arrayrulecolor{black}}
\newcommand{\catlab}[4]{\rotatebox[origin=l]{90}{\textcolor{gInk#1}{\textbf{\small #2}~{\scriptsize $N$\,=\,#4, $n$\,=\,#3}}}}
\newcommand{\gHeader}{%
  \toprule
  & \begin{tabular}{Q}
    \textbf{\small Guideline} &
    \textbf{\small C / R / A}\newline{\scriptsize(\% Yes)} &
    \textbf{\small How participants described operationalizing it in practice}
    \end{tabular}\\
  \midrule}

\begin{table*}[p!]
\caption{(a) Guidelines for responsibly integrating AI in SE work, with \textbf{C}larity, \textbf{R}elevance, and \textbf{A}doption ratings by our study's developers, and ways they described operationalizing each guideline. ($N$: Phase~1 participants whose responses the category's guidelines were synthesized from; $n$: Phase~2 participants who rated them). \emph{Unsure} responses (6\% of all ratings) are reported in the supplement~\cite{supplemental}. Lighter bars: below the 70\% threshold (vertical~line); $^\star$ guidelines: adoption below 70\%.}
\label{tab:guidelines-scenarios}
\small
\setlength{\tabcolsep}{3pt}
\renewcommand{\arraystretch}{0.93}
\begin{tabular}{@{}>{\centering\arraybackslash}m{\gCatW}@{\hspace{3pt}}l@{}}
\gHeader
\gtint{DEV}\catlab{DEV}{Development}{226}{816} &
\begin{tabular}{Q}
\gtintrow{DEV}\grow{\textcolor{gKwDEV}{DV1}:}{\textcolor{gKwDEV}{\textsc{\textbf{Use}}} AI to automate routine, verifiable development tasks under human oversight; \textcolor{gKwDEV}{\textsc{\textbf{demand}}} AI to produce reviewable steps and plan of~action.}{96}{98}{97}{Automate recurring upkeep such as log parsing, CI, and data updates; Delegate mechanical refactors that you dictate and can verify; Encode repeated work as checked-in skills, instructions, or custom agents; Orchestrate specialized agents that hand off work and request clarification; Have AI refine plans before authorizing implementation; Have AI give pull requests a first review while retaining~human~approval}
\gsep
\gtintrow{DEV}\grow{\textcolor{gKwDEV}{DV2}:}{\textcolor{gKwDEV}{\textsc{\textbf{Use}}} AI to find and synthesize context from multiple sources to make changes with expected trade-offs; \textcolor{gKwDEV}{\textsc{\textbf{demand}}} AI to follow standards and disclose scope and~assumptions.}{91}{97}{95}{Correlate dashboards, logs, and incidents to guide engineering decisions; Trace change impacts across repositories and related applications; Maintain indexed project context that agents can draw on; Constrain agents with documented rules and responsibility boundaries; Compare design options with AI, then choose the direction yourself; Make AI state scope, assumptions, and~confidence~explicitly}
\gsep
\gtintrow{DEV}\grow{\textcolor{gKwDEV}{DV3}:}{\textcolor{gKwDEV}{\textsc{\textbf{Use}}} AI to make sense of unfamiliar project context and diagnose problems; \textcolor{gKwDEV}{\textsc{\textbf{demand}}} AI to provide traceable, source-linked explanations and~recommendations.}{98}{98}{98}{Navigate unfamiliar code with AI before changing it; Demand explicit sources for every AI claim; Have AI investigate incidents and locate affected code paths; Condense dense documentation into usable explanations; Create and maintain AI-written explainer documents for complex systems; Ask AI to pinpoint logic flaws and error-handling problems}
\gsep
\gtintrow{DEV}\grow{\textcolor{gKwDEV}{DV4}:}{\textcolor{gKwDEV}{\textsc{\textbf{Use}}} AI to create changes in human-reviewable increments; \textcolor{gKwDEV}{\textsc{\textbf{demand}}} AI to hold consequential changes for human sign-off before applying~them.}{92}{98}{94}{Break large changes into small increments and verify each one; Let AI diagnose and propose changes in small, reviewable steps, and verify its sources and assumptions; Review generated code continuously and redirect the agent during implementation; Let AI reach a draft pull request before intervening; Run an independent review agent before humans inspect the pull request; Block consequential actions until a human explicitly~approves~them}
\gsep
\gtintrow{DEV}\grow{\textcolor{gKwDEV}{DV5}:}{\textcolor{gKwDEV}{\textsc{\textbf{Use}}} AI as a coach for skill building; \textcolor{gKwDEV}{\textsc{\textbf{demand}}} AI to provide clear explanations and uncertainty~indicators.}{94}{89}{88}{Learn by solving real tasks with AI; Talk through ideas with AI as a pair-programming partner; Use AI coaching to prepare~junior~developers}
\gsep
\gtintrow{DEV}\grow{\textcolor{gKwDEV}{DV6}:}{\textcolor{gKwDEV}{\textsc{\textbf{Prohibit}}} AI from making or finalizing security- and performance-critical development~decisions.}{90}{91}{84}{Reserve final security- and performance-critical decisions for human sign-off, and approve sensitive~changes~personally}
\end{tabular}\\
\midrule
\gtint{DNP}\catlab{DNP}{Design \& Planning}{158}{548} &
\begin{tabular}{Q}
\gtintrow{DNP}\grow{\textcolor{gKwDNP}{DP1}:}{\textcolor{gKwDNP}{\textsc{\textbf{Use}}} AI to find and synthesize information from different sources with clear source attribution; \textcolor{gKwDNP}{\textsc{\textbf{demand}}} AI to surface historical context and inconsistencies for human~review.}{96}{94}{95}{Consolidate internal project sources before planning or designing; Research prior solutions and other teams' practices before designing; Demand that AI link its findings to specific sources so humans can verify them; Have AI summarize requirements after~client~meetings}
\gsep
\gtintrow{DNP}\grow{\textcolor{gKwDNP}{DP2}:}{\textcolor{gKwDNP}{\textsc{\textbf{Use}}} AI to explore contextually grounded design and architecture alternatives with expected trade-off and impact analysis; \textcolor{gKwDNP}{\textsc{\textbf{demand}}} AI to maintain provenance of design~decisions.}{93}{95}{95}{Brainstorm architecture ideas with AI before development starts; Generate several design options with trade-offs, then choose among them; Have AI draft the design, then refine it with expert review; Have AI critique a draft design for gaps, unstated assumptions, and~corner~cases}
\gsep
\gtintrow{DNP}\grow{\textcolor{gKwDNP}{DP3}:}{\textcolor{gKwDNP}{\textsc{\textbf{Use}}} AI to convert reviewed designs into downstream artifacts; \textcolor{gKwDNP}{\textsc{\textbf{demand}}} AI to make every artifact traceable to its reviewed~sources.}{94}{92}{94}{Generate work items and implementation plans from approved designs; Draft review-ready specifications and design documents from requirements; Approve designs before AI converts them into downstream artifacts; Keep generated artifacts linked to source designs across teams~and~time}
\gsep
\gtintrow{DNP}\grow{\textcolor{gKwDNP}{DP4}:}{\textcolor{gKwDNP}{\textsc{\textbf{Use}}} AI to inform task plans, schedules, and resource assignments fairly to fit the project's context and people's interests and skills; \textcolor{gKwDNP}{\textsc{\textbf{demand}}} AI to disclose its assumptions and uncertainty~explicitly.}{92}{78}{79}{Set constraints, let AI draft the plan, then review and edit it; Run project management with AI across the full delivery cycle; Gate implementation on plan confidence or~human~review}
\gsep
\gtintrow{DNP}\grow{\textcolor{gKwDNP}{DP5}:}{\textcolor{gKwDNP}{\textsc{\textbf{Use}}} AI to automate routine work-item creation and updates (e.g., status reports); \textcolor{gKwDNP}{\textsc{\textbf{demand}}} AI to keep every change easy to review and~undo.}{94}{95}{92}{Automate work items, status updates, and board bookkeeping; Track status, next steps, blockers, reminders, and~rollout~gates}
\gsep
\gtintrow{DNP}\grow{\textcolor{gKwDNP}{DP6}:}{\textcolor{gKwDNP}{\textsc{\textbf{Prohibit}}} AI from making or finalizing consequential design and planning decisions, such as inferring unstated requirements, deciding on architecture or business logic, or automatically assigning~tasks.}{90}{92}{84}{Have AI surface assumptions and rejected alternatives for confirmation before proceeding; Keep final design and planning decisions behind~human~sign-off}
\gsep
\gtintrow{DNP}\grow{\textcolor{gKwDNP}{DP7}:}{\textcolor{gKwDNP}{\textsc{\textbf{Limit}}} AI's role in strategic vision setting, stakeholder alignment, and human-relational~work.}{85}{71}{71}{Bring AI in to manage the work only after people have set the strategic direction and decided what~to~build}
\end{tabular}\\
\midrule
\gtint{QRM}\catlab{QRM}{Quality \& Risk}{128}{401} &
\begin{tabular}{Q}
\gtintrow{QRM}\grow{\textcolor{gKwQRM}{QR1}:}{\textcolor{gKwQRM}{\textsc{\textbf{Use}}} AI for continuous, proactive QA that understands the project's context; \textcolor{gKwQRM}{\textsc{\textbf{demand}}} AI to surface findings ranked by risk, linked to evidence, and paired with reviewable fix~guidance.}{91}{92}{94}{Continuously scan code, telemetry, and releases to catch emerging risks; Demand that AI rank its findings by risk to focus limited review effort; Investigate logs and failures with AI to reproduce issues and propose fixes; Demand evidence and citations behind every AI quality finding; Assess testing gaps from code, change, and coverage data; Use AI during pull-request review to flag risks~and~gaps}
\gsep
\gtintrow{QRM}\grow{\textcolor{gKwQRM}{QR2}:}{\textcolor{gKwQRM}{\textsc{\textbf{Use}}} AI to automate routine QA and compliance tasks; \textcolor{gKwQRM}{\textsc{\textbf{demand}}} AI to provide uncertainty indicators on every~output.}{86}{88}{90}{Automate recurring compliance checks, alerts, and result routing; Assign neglected QA backlogs to AI when teams lack capacity; Run end-to-end tests through browser automation and test infrastructure; Have AI write tests while developers retain~testing~ownership}
\gsep
\gtintrow{QRM}\grow{\textcolor{gKwQRM}{QR3}:}{\textcolor{gKwQRM}{\textsc{\textbf{Limit}}} AI's handling of personally identifiable and sensitive compliance data to least-privilege~access.}{89}{67}{79}{Make clear which data and tools AI may use, and scope its access to sensitive data}
\gsep
\gtintrow{QRM}\grow{\textcolor{gKwQRM}{QR4}:}{\textcolor{gKwQRM}{\textsc{\textbf{Prohibit}}} AI from self-certifying the correctness or quality of its own~output.}{89}{91}{83}{Gate AI output with deterministic tests; Use automated builds to catch false claims in AI's reports of success; Check that AI-written tests exercise the intended behavior; Use a separate agent as a reviewer to check AI's contributions; Keep approval of AI findings, fixes, and commits with~a~human}
\end{tabular}\\
\bottomrule
\end{tabular}
\end{table*}

\begin{table*}[t!]
\addtocounter{table}{-1}
\caption{(b) Guidelines for responsibly integrating AI in SE work (continued).}
\small
\setlength{\tabcolsep}{3pt}
\renewcommand{\arraystretch}{0.93}
\begin{tabular}{@{}>{\centering\arraybackslash}m{\gCatW}@{\hspace{3pt}}l@{}}
\gHeader
\gtint{INF}\catlab{INF}{Infrastructure \& Ops}{84}{283} &
\begin{tabular}{Q}
\gtintrow{INF}\grow{\textcolor{gKwINF}{IO1}:}{\textcolor{gKwINF}{\textsc{\textbf{Use}}} AI for continuous, telemetry-driven DevOps monitoring, analysis, and release-readiness assessment; \textcolor{gKwINF}{\textsc{\textbf{demand}}} AI to provide review-ready summaries and pre-incident~warnings.}{93}{90}{96}{Continuously monitor telemetry for anomalies and emerging failures; Correlate logs and telemetry to diagnose incidents and surface likely causes; Triage every incoming incident before human handling; Watch deployments and releases through completion; Triage failed tests and assess release quality; Report patterns and trends across~support~incidents}
\gsep
\gtintrow{INF}\grow{\textcolor{gKwINF}{IO2}:}{\textcolor{gKwINF}{\textsc{\textbf{Use}}} AI to generate Infrastructure-as-Code and configuration workflows from intent; \textcolor{gKwINF}{\textsc{\textbf{demand}}} AI to provide a human-readable diff before executing~changes.}{93}{89}{94}{Generate infrastructure-as-code from high-level intent; Demand a reviewable diff or plan before infrastructure changes run; Build and maintain CI/CD pipelines from~stated~requirements}
\gsep
\gtintrow{INF}\grow{\textcolor{gKwINF}{IO3}:}{\textcolor{gKwINF}{\textsc{\textbf{Use}}} AI to automate routine operations tasks; \textcolor{gKwINF}{\textsc{\textbf{demand}}} AI to provide uncertainty indicators and clear, traceable explanations on every~recommendation.}{92}{96}{97}{Draft remediation steps, fixes, and runbooks for engineers; Convert remaining manual operations into supervised scripts and workflows; Wire SRE agents to incident triggers, telemetry, and scoped action tools; Require reviewable summaries, confidence, and explanations before letting AI make~production~changes}
\gsep
\gtintrow{INF}\grow{\textcolor{gKwINF}{IO4}:}{\textcolor{gKwINF}{\textsc{\textbf{Prohibit}}} AI from modifying production environments, critical infrastructure, security configurations, or access controls on its~own.}{92}{91}{86}{Fence AI with firewalls, sandboxes, and identity controls while it helps run infrastructure; Hold consequential environment changes for human approval~or~execution}
\gsep
\gtintrow{INF}\grow{\textcolor{gKwINF}{IO5}:}{\textcolor{gKwINF}{\textsc{\textbf{Prohibit}}} AI from autonomously communicating with stakeholders or making emergency-response~decisions.}{88}{81}{84}{Assign a person accountable for emergency decisions and messages sent under their name: route emergency decisions to a human, check messages before they are sent, and let agents report only routine release status}
\end{tabular}\\
\midrule
\gtint{MW}\catlab{MW}{Meta-work}{90}{532} &
\begin{tabular}{Q}
\gtintrow{MW}\grow{\textcolor{gKwMW}{MW1}:}{\textcolor{gKwMW}{\textsc{\textbf{Use}}} AI to draft stakeholder communications adapted to each audience's needs and intended call to action; \textcolor{gKwMW}{\textsc{\textbf{demand}}} AI to provide drafts traceable to their source~materials.}{89}{67}{71}{Draft routine communication from material developers supply, and keep interpersonal messages their own: AI drafts newsletters from recent updates and summarizes them, while developers write to~colleagues~themselves}
\gsep
\gtintrow{MW}\grow{\textcolor{gKwMW}{MW2}$^\star$:}{\textcolor{gKwMW}{\textsc{\textbf{Use}}} AI to automate routine communication and collaboration logistics; \textcolor{gKwMW}{\textsc{\textbf{demand}}} AI to never dispatch consequential stakeholder-facing messages without~approval.}{93}{71}{62}{Track work and compile stakeholder updates from Azure DevOps and Teams; Draft release notes and routine announcements from commits, then review~before~sending}
\gsep
\gtintrow{MW}\grow{\textcolor{gKwMW}{MW3}:}{\textcolor{gKwMW}{\textsc{\textbf{Use}}} AI to adapt learning to knowledge level and project needs; \textcolor{gKwMW}{\textsc{\textbf{demand}}} AI to scaffold active, hands-on practice and deliberate~thinking.}{83}{85}{87}{Summarize and filter trusted learning resources around the learner's question; Have AI explain each code line and write your own~learning~notes}
\gsep
\gtintrow{MW}\grow{\textcolor{gKwMW}{MW4}:}{\textcolor{gKwMW}{\textsc{\textbf{Use}}} AI as a thought partner to challenge and refine ideas; \textcolor{gKwMW}{\textsc{\textbf{demand}}} AI to defer final decisions to human~judgment.}{94}{94}{90}{Bring ideas to AI for questions, challenge, and refinement; Set up dedicated AI projects and personalize settings for thinking through ideas; Use AI to help find/summarize resources}
\gsep
\gtintrow{MW}\grow{\textcolor{gKwMW}{MW5}:}{\textcolor{gKwMW}{\textsc{\textbf{Use}}} AI to curate, validate, and maintain documentation and shared project context; \textcolor{gKwMW}{\textsc{\textbf{demand}}} AI to preserve fidelity to its sources and flag stale or conflicting context for human~review.}{96}{93}{94}{Update and validate shared documentation as work evolves; Scan documentation against code and changes to~flag~drift}
\gsep
\gtintrow{MW}\grow{\textcolor{gKwMW}{MW6}:}{\textcolor{gKwMW}{\textsc{\textbf{Use}}} AI to create personalized training plans and materials for onboarding and learning new technologies; \textcolor{gKwMW}{\textsc{\textbf{demand}}} AI to provide editable, source-linked~drafts.}{94}{77}{78}{Generate up-to-date training and onboarding materials: wiki pages for knowledge transfer, guides to new AI tools adapted to the learner's skill, and overviews of features owned by~other~teams}
\gsep
\gtintrow{MW}\grow{\textcolor{gKwMW}{MW7}$^\star$:}{\textcolor{gKwMW}{\textsc{\textbf{Limit}}} AI's role in mentoring to preserve mentor and mentee development and tacit knowledge~transfer.}{83}{64}{61}{Keep mentoring and people decisions human: face-to-face time and whiteboarding with new hires, messages to colleagues written by people, and AI brought in afterward for implementation and checking}
\end{tabular}\\
\midrule
\gtint{XC}\catlab{XC}{Cross-cutting}{190}{860} &
\begin{tabular}{Q}
\gtintrow{XC}\grow{\textcolor{gKwXC}{CC1}:}{\textcolor{gKwXC}{\textsc{\textbf{Demand}}} AI to ask for clarification instead of acting on missing or inferred~context.}{97}{99}{98}{Make AI ask before ambiguous work begins: instruct it to ask about intent or open choices rather than guess, surface unknowns and assumptions for confirmation, and clarify feedback before acting~on~it}
\gsep
\gtintrow{XC}\grow{\textcolor{gKwXC}{CC2}:}{\textcolor{gKwXC}{\textsc{\textbf{Demand}}} AI to prefer simple, deliberate solutions over superficial fixes or unnecessarily complex~ones.}{94}{98}{94}{Repeatedly trim unnecessarily complex code and verbose documentation; Choose the leanest complete root-cause fix; Pull AI back from over-broad changes; Expect AI to avoid unnecessary complexity without~being~told}
\gsep
\gtintrow{XC}\grow{\textcolor{gKwXC}{CC3}:}{\textcolor{gKwXC}{\textsc{\textbf{Demand}}} AI to safeguard the privacy and security of the data it handles, exposing or retaining it only as the task~requires.}{93}{91}{93}{Make AI apply security and privacy rules; Keep confidential information outside the repository so AI cannot~access~it}
\gsep
\gtintrow{XC}\grow{\textcolor{gKwXC}{CC4}:}{\textcolor{gKwXC}{\textsc{\textbf{Demand}}} AI to back its claims with cited, verifiable evidence, and to clearly state its assumptions, reasoning, scope, and~uncertainty.}{96}{99}{99}{Cite specific code locations and source documents; State reasoning, assumptions, uncertainty, and missing scenarios; Codify evidence checks as standing~agent~rules}
\gsep
\gtintrow{XC}\grow{\textcolor{gKwXC}{CC5}:}{\textcolor{gKwXC}{\textsc{\textbf{Demand}}} AI to deliver its output in atomic, self-contained pieces a human can~review.}{93}{98}{92}{Require AI to deliver atomic, reviewable changes; Decompose work yourself and verify each step; Give AI a guide with examples people have checked for easy review; Review project plans against requirements and add notes about the whole plan; Split AI output into pieces so team members can~share~review}
\gsep
\gtintrow{XC}\grow{\textcolor{gKwXC}{CC6}:}{\textcolor{gKwXC}{\textsc{\textbf{Demand}}} AI to clearly annotate its generated contributions and communications to distinguish them from human-authored~content.}{93}{95}{81}{Mark AI's contributions where reviewers see them: label AI-written review comments, mark generated content as AI-generated and human-reviewed, and credit AI as co-author on~accepted~code}
\end{tabular}\\
\bottomrule
\end{tabular}
\end{table*}

%% file: sec/4.Results-new.tex
\section{Results}
\label{sec:results}
Overall, developers in our study strongly endorsed the guidelines: they found them clearly worded (\textbf{C}larity 92\%), relevant to their SE work (\textbf{R}elevance 91\%), and worth adopting (\textbf{A}doption 89\%), and frequently described them as practices they were already working toward. Table~\ref{tab:guidelines-scenarios} reports these guidelines (29 across five SE task categories and 6 cross-cutting ones), with participants’ clarity, relevance, and adoption ratings of each and the ways they operationalized these guidelines in their daily work.

Participants' ratings did not differ significantly between domains, so we report domain-level differences only where they were substantive, as for Meta-work (Sec.~\ref{subsec:res-mw}).

\subsection[Development]{Development (N=816, n=226; \textbf{C} 94\%, \textbf{R} 95\%, \textbf{A} 93\%)}
\label{subsec:res-dev}
The Development category drew Phase~2 participants' (n=226) strongest endorsement among the five SE task categories (\textbf{R} 95\%, \textbf{A} 93\%; Table~\ref{tab:guidelines-scenarios}).
Recall that these percentages rate relevance and usefulness to participants' own work (\textbf{R}elevance: \enquote{\ldots applies to \textit{my} work}; \textbf{A}doption: \enquote{I want to adopt this in \textit{my} work}).
In the Development category, Phase~1 participants  (N=816) made three central demands for responsibly integrating AI into their own work:

\textbf{(1)~When AI helped participants understand code and context, that its assumptions and explanations be explicit and traceable:}
Participants in both phases described using AI to understand unfamiliar code and diagnose problems, \textit{navigating unfamiliar code with AI before changing it} and \textit{condensing dense documentation into usable explanations},
while \textit{demanding explicit sources for every AI claim} (\gl{DV}{3}; R 98\%, A 98\%):
\pq{I often need to tell AI \enquote{show me the source behind claim X} [so] I know where to look in the code} (MS24).

They further noted using AI to synthesize context from multiple sources before making changes, for example, \textit{correlating dashboards, logs, and incidents to guide engineering decisions}, while requiring AI to \textit{state scope, assumptions, and confidence explicitly} (\gl{DV}{2}; R 97\%, A 95\%): \pq{\ldots having AI synthesize context\ldots and explain it with traceable, source-linked reasoning\ldots disclose its assumptions and flag uncertainty instead of sounding confident when it's guessing} (MS45).

This demand was also echoed when participants used AI for skill building, where they would \textit{learn by solving real tasks with AI} or \textit{talk through ideas with AI as a pair-programming partner}, while demanding clear explanations and indicators of its uncertainty (\gl{DV}{5}; R 89\%, A 88\%): \pq{It has been a great \enquote{sounding board} for me to bounce ideas off of before committing to any given direction} (MS158).

\textbf{(2)~When AI did development work, that its contributions arrive as plans and increments participants could effectively review:}
When AI implemented code changes or automated routine development tasks, participants required it to first propose a plan of action they could review (\gl{DV}{1}; R 98\%, A 97\%), and then to deliver its changes as self-contained, atomic units that could each be reviewed on its own (\gl{DV}{4}; R 98\%, A 94\%).
MS28 described the desired sequence as: \pq{AI helps me to develop the software architecture\ldots and formulate an implementation plan. Once the plan is rock-solid, I ask the AI to implement the code changes, and then I verify the implementation}.
Others kept AI's changes small, as in \pq{automate routine changes in small, reviewable steps}~(MS117).

\textbf{(3)~Before AI's changes took effect, that consequential decisions wait for human sign-off:}
Participants kept consequential changes for their own approval (\gl{DV}{4}) and wanted that approval enforced through \pq{intentional human gates, enforced by actual governance tools (hooks, vs prompts)} (MS179).
They would also \textit{constrain agents with documented rules and responsibility boundaries} (\gl{DV}{2}), for example, by \pq{\ldots injecting those rules into our agents (including Azure best practices for instance)} (MS142).\looseness=-1

Especially with security- and performance-critical work, participants chose to \textit{reserve final security- and performance-critical decisions for human sign-off} (\gl{DV}{6}; R 91\%, A 84\%): \pq{For\ldots performance, sec[urity] decisions, it is good to brainstorm with AI and figure out the options\ldots and let the human make the final decisions} (MS110).

The Development category guidelines' emphasis on the importance of reviewing AI contributions raises the question of whether humans can effectively review AI contributions at their current scale (e.g.,~\cite{isaac2026nytimes-codeReview}).
Some participants also pointed this out and suggested mitigations such as this one:
\pq{I have it do automated reviews with second models on PRs created, before my team has an opportunity to review} (MS126).
Whether this mitigation will be enough given the scale of this issue is still an open question.\looseness=-1


\subsection[Design \& Planning]{Design \& Planning (N=548, n=158; \textbf{C} 92\%, \textbf{R} 89\%, \textbf{A} 88\%)}
\label{subsec:res-dnp}


Phase~2 participants also strongly endorsed the Design \& Planning guidelines (\textbf{R} 89\%, \textbf{A} 88\%), which at their core captured three demands that Phase~1 participants made of AI:

\textbf{(1)~When AI helped participants research and explore designs, that its sources be attributed and design decisions keep their provenance:}
Participants described using AI to \textit{consolidate internal project sources} and \textit{research prior solutions and other teams' practices before designing}, while \textit{demanding that AI link its findings to specific sources so humans can verify them} (\gl{DP}{1}; R 94\%, A 95\%).
PF24 described how their team built this into its workflow: \pq{We use AI to ingest information from multiple sources, and curate it. Every curated claim links to one or more raw source documents}.

The same demand held when they would \textit{generate several design options with trade-offs, then choose among them}, and \textit{have AI critique a draft design for gaps, unstated assumptions, and corner cases}---always keeping the reasoning behind each design decision explicit (\gl{DP}{2}; R 95\%, A 95\%).

\textbf{(2)~When AI turned approved designs into artifacts and work items, that each stay traceable to the design and easy to review:}
Using AI to \textit{generate work items and implementation plans from approved designs} and \textit{automate work items, status updates, and board bookkeeping}, participants required artifacts to trace to their reviewed design (\gl{DP}{3}; R 92\%, A 94\%) such that each change would be easy to review and undo (\gl{DP}{5}; R 95\%, A 92\%): \pq{I use AI to generate \ldots design documents, status updates\ldots and implementation plans, while ensuring the outputs remain traceable and are reviewed by humans \ldots}~(MS151).

\textbf{(3)~Throughout, that consequential design and planning decisions retain human sign-off:}
Such decisions included inferring unstated requirements, deciding on architecture or business logic, and assigning tasks~(\gl{DP}{6};~R~92\%,~A~84\%).

A few, however, were reluctant to adopt \gl{DP}{6}, wanting AI to make more of these decisions: \pq{Historically these functionary roles seemed like \enquote{leadership} but with AI it's starting to be the creative/authoring traits that help me excel, while logistical decisions can be made by AI much better} (MS41).

Notably, endorsement in this category was relatively lower for the two guidelines that centered on relationship-centric work: using AI to inform schedules and resource assignments (\gl{DP}{4}; R 78\%, A 79\%) and limiting AI's role in strategic vision, stakeholder alignment, and human-relational work (\gl{DP}{7}; R 71\%, A 71\%), although a majority ($>$70\%) still endorsed both.

Breaking these differences down by participants' demographics, we found that senior developers in Phase~2 rated \gl{DP}{4} relatively less relevant to their work (71\% vs.\ 85\% of junior and mid-career developers).
\gl{DP}{7}, on the other hand, was slightly less relevant to junior and mid-career developers (70\% vs.\ 72\% of senior developers).
Some of them noted that strategy and stakeholder alignment were not yet a big part of their daily work.
Furthermore, participants with more AI experience were less willing to adopt \gl{DP}{7} in their work (70\% vs.\ 74.3\% of those with low to moderate AI experience).


\subsection[Quality \& Risk]{Quality \& Risk (N=401, n=128; \textbf{C} 89\%, \textbf{R} 84\%, \textbf{A} 87\%)}
\label{subsec:res-qrm}

Phase~2 participants broadly endorsed the Quality \& Risk guidelines (\textbf{R} 84\%, \textbf{A} 87\%).
The central theme of these guidelines was what they demanded of AI as it extended testing and risk scanning beyond what their teams could check:

\textbf{(1)~When AI scanned for risks at scale, that every finding be traceable to its evidence:}
Participants described using AI to \textit{continuously scan code, telemetry, and releases to catch emerging risks} and \textit{investigate logs and failures with AI to reproduce issues and propose fixes}, while \textit{demanding that AI rank its findings by risk to focus limited review effort} and \textit{demanding evidence and citations behind every AI quality finding} (\gl{QR}{1}; R 92\%, A 94\%).
MS8 described why this mattered: \pq{\ldots we've already seen\ldots hallucinated security reports can overwhelm the systems' ability to deal with them\ldots there's an asymmetry where creating this stuff is vastly cheaper than verifying that it's real}.

Similarly, for routine QA and compliance, participants described \textit{automating recurring compliance checks, alerts, and result routing} and \textit{having AI write tests while developers retain testing ownership}, while demanding that AI indicate its uncertainty explicitly (\gl{QR}{2}; R 88\%, A 90\%).

\textbf{(2)~When AI handled sensitive data, that it be given least-privilege access:}
67\% of participants said limiting AI's handling of personally identifiable and sensitive data was relevant to their work; still, a larger share (79\%) wanted to adopt it in their work (\gl{QR}{3}).
Participants further noted that team policies should \textit{make clear which data and tools AI may use, and scope its access to sensitive data}; e.g., \pq{access should be granted in expanding circles\ldots start with a limited scope\ldots If not, expand the scope and repeat}~(MS139).

\textbf{(3)~Before participants accepted AI's output, that its work be verified independently:}
Participants strongly endorsed prohibiting AI from certifying its own output (\gl{QR}{4}; R 91\%, A 83\%), and would \textit{use a separate agent as a reviewer to check AI's contributions}: \pq{subagents with clean context and from different family of models to review the output of the other to be more unbiased} (PF15).
Echoing the Development category, some participants noted that AI-based verification would need to grow alongside human review: \pq{Humans are not going to be able to keep up \ldots The focus should be on making good evals that can be used to gatekeep AI} (MS190).


\subsection[Infrastructure \& Ops]{Infrastructure \& Ops (N=283, n=84; \textbf{C} 91\%, \textbf{R} 89\%, \textbf{A} 92\%)}
\label{subsec:res-inf}

Phase~2 participants strongly endorsed the Infrastructure \& Ops guidelines (\textbf{R} 89\%, \textbf{A} 92\%).
Phase~1 participants required three core behaviors of AI in the operations lifecycle:

\textbf{(1)~When AI monitored systems, that its summaries be review-ready and state its confidence:}
Participants were especially keen to adopt continuous, telemetry-driven monitoring (\gl{IO}{1}; R 90\%, A 96\%), using AI to \textit{continuously monitor telemetry for anomalies and emerging failures} and \textit{correlate logs and telemetry to diagnose incidents and surface likely causes}, while demanding summaries they could effectively review, with AI's confidence clearly stated: \pq{What it misses often is the interpretation, so review-ready summaries with confidence level clearly described are very useful} (MS125).

\textbf{(2)~When AI changed infrastructure, that every change come with a human-readable diff or plan:}
Participants described using AI to \textit{generate infrastructure-as-code from high-level intent} and \textit{build and maintain CI/CD pipelines from requirements}, while \textit{demanding a reviewable diff or plan before infrastructure changes run} (\gl{IO}{2}; R 89\%, A 94\%): \pq{Using AI with infrastructure as code like Terraform is helpful, basically acting as one level higher of the declarative intent for infra as code\ldots It's dangerous to blindly trust what it says\ldots A real staging environment is crucial to validate these things} (MS47).

For routine operations, they would \textit{draft remediation steps, fixes, and runbooks for engineers}, while demanding provenance and AI's explanations before letting it make production changes (\gl{IO}{3}; R 96\%, A 97\%): \pq{I'd use AI to analyze monitoring data, summarize incidents, identify likely root causes, and suggest remediation steps\ldots Recommendations should be explainable, auditable, and require approval\ldots}~(MS88).

\textbf{(3)~In production and emergencies, that a human decide and stay accountable:}
Participants gated AI from changing production on its own (\gl{IO}{4}; R 91\%, A 86\%), \textit{holding consequential environment changes for human approval or execution} and \textit{fencing AI with firewalls, sandboxes, and identity controls while it helps run infrastructure}.
As MS3 put it: \pq{I am already telling AI to help me build scripts which can run manual operations \ldots on production systems. I don't necessarily think AI should be running these commands itself}.

One participant, however, believed agents should be able to fix routine production issues: \pq{We need a way to have agents with identities and access to prod resources while making sure a human is still in control\ldots} (MS67).\looseness=-1

In emergencies and high-stakes communication (\gl{IO}{5}; R 81\%, A 84\%), participants would \textit{assign a person accountable for emergency decisions and messages sent under their name}, noting that \pq{a computer cannot be held accountable; therefore, a computer should never make a management decision} (MS80).

\subsection[Meta-work]{Meta-work (N=532, n=90; \textbf{C} 90\%, \textbf{R} 79\%, \textbf{A} 78\%)}
\label{subsec:res-mw}

Phase~2 participants endorsed the Meta-work guidelines (\textbf{R} 79\%, \textbf{A} 78\%), though less uniformly than those in the other categories.
This work spanned learning, thinking through ideas, documentation, communication, and mentoring, and Phase~1 participants' demands of AI depended on whom the work involved: the more it involved other people, the less they wanted AI acting on their behalf:

\textbf{(1)~When participants built knowledge for themselves, that AI scaffold active learning and leave the final judgment to them:}
Participants described using AI to adapt learning to their needs, \textit{summarizing and filtering trusted learning resources around the learner's question}, while demanding active, hands-on practice and deliberate thinking (\gl{MW}{3}; R 85\%, A 87\%): \pq{Always have AI explain outputs not only for validation but for learning purposes, making sure I understand each line of code and writing notes for my own understanding, separate from project goals} (PF23).

With AI as a thought partner, they would \textit{bring ideas to AI for questions, challenge, and refinement}, while keeping the final judgment for themselves (\gl{MW}{4}; R 94\%, A 90\%): \pq{I've used it to push back on my ideas\ldots Sometimes it brings up good points I didn't think of before} (MS197).

\textbf{(2)~When participants communicated with others, that their voice stay their own and no consequential message go out without their sign-off:}
A majority endorsed drafting stakeholder communications traceable to their source material (\gl{MW}{1}; R 67\%, A 71\%) and automating routine communication with sign-off on consequential messages (\gl{MW}{2}; R 71\%, A 62\%), although \gl{MW}{2}'s adoption dipped below 70\%.
Those who adopted them would \textit{draft routine communication from material developers supply, and keep interpersonal messages their own} and \textit{track work and compile stakeholder updates from Azure DevOps and Teams}: \pq{Drafting documents for review is valuable...directly communicating without review is too risky and is liable to cause more problems than it solves} (MS169).

Those who were reluctant wanted to keep AI out of communication entirely: \pq{I don't want AI to be my voice, my voice is part of what makes me who I am\ldots Communication is a skill that if we offload it we will not get better at it} (MS90).
Both guidelines were more relevant to participants with more AI experience (74\% vs.\ 62\%).
Some found value in handing routine communication, such as release notes and newsletters, to AI while retaining sign-off on it.

\textbf{(3)~In mentoring, that humans remain the mentors:}
Participants described using AI to \textit{generate up-to-date training and onboarding materials} (\gl{MW}{6}; R 77\%, A 78\%) while limiting AI's role in mentoring (\gl{MW}{7}; R 64\%, A 61\%).
Interestingly, \gl{MW}{7} drew more reluctance than we expected: about a third of participants did not endorse it, reflecting that mentoring rests more on individual choice and each participant's principles about AI.
Those who wanted to adopt it would \textit{keep mentoring and people decisions human}: \pq{When I mentored a new hire, it was helpful to the mentee to have face-to-face time and whiteboarding sessions, then later use the AI for implementation and extra layer for checking} (MS155).
Moreover, juniors found \gl{MW}{7} more relevant than seniors (85\% vs.\ 76\%), noting the value of the human element in mentoring.

The Meta-work category also showed a difference between the domains: Pfizer participants were more reluctant than their Microsoft peers to adopt AI for routine communication, and even more so for AI-generated training material (pooled across Meta-work: 71\% vs.\ 80\%).
Some drew the line at AI-written communication altogether: \pq{Communications should be filled with intent and decisions. I don't want to read AI-generated text, and I've noticed no one else reads it either} (PF20).

\subsection[Cross-cutting]{Cross-cutting (N=860, n=190; \textbf{C} 94\%, \textbf{R} 97\%, \textbf{A} 93\%)}
\label{subsec:res-xc}

Overall, the cross-cutting guidelines drew the highest ratings across all categories from Phase~2 participants (\textbf{R} 97\%, \textbf{A} 93\%).
The essence of these guidelines, contributed by the Phase~1 participants, lay in the conditions they placed on AI's work in every category, in addition to the category-level ones described above.
These demands followed AI's work end-to-end.
Finally, clearly annotating AI's contributions, which concerns who gets credit for the work, interestingly divided participants.

\textbf{(1)~Before AI acted, that it ask when in doubt instead of making assumptions:}
Participants would \textit{make AI ask before ambiguous work begins} (\gl{CC}{1}; R 99\%, A 98\%), having it surface unknowns and assumptions for confirmation first: \pq{I have a few sub agents/skills where I clearly state for the agent to surface unknowns and assumptions. Then to get input from the user on those before performing any next steps} (MS14).

One participant offered nuance on deciding when to ask: \pq{clarification should be risk-based \ldots Excessive questions can interrupt work, while silent assumptions on consequential decisions create unacceptable risk} (MS76).

\textbf{(2)~While AI worked, that it choose deliberate solutions and protect data:}
Participants demanded that AI prefer simple over unnecessarily complex solutions and deliberate fixes over \enquote{band-aid} ones (\gl{CC}{2}; R 98\%, A 94\%): \pq{AI [tends] to generate very complex, WORKING fixes, that are BAD because they diverge from the intended design} (MS7).

Some participants, meanwhile, cautioned that simplicity alone was not enough; a solution also had to be deliberate: \pq{I have seen plenty of simple, but wrong solutions. And sometimes the right solution requires complexity} (MS123).


To protect data, they would \textit{make AI apply security and privacy rules} or \textit{keep confidential information outside the repository to prevent AI access} (\gl{CC}{3}; R 91\%, A 93\%): \pq{I have to direct AI to honor security best practices. If AI knew and applied these rules, that'd be a huge advancement} (MS111).

\textbf{(3)~When AI handed back work, that its claims carry evidence and its output be in self-contained, human-reviewable units:}
Participants demanded that AI \textit{cite code locations and source documents} and \textit{state reasoning, assumptions, uncertainty, and missing scenarios} (\gl{CC}{4}; R 99\%, A 99\%): \pq{One of [my] largest frustrations\ldots constantly having to remind [AI] to 
[base its work] on the actual sources}~(MS200).

They also required AI to \textit{deliver atomic, reviewable changes} (\gl{CC}{5}; R 98\%, A 92\%): \pq{I try to decompose my tasks into digestible atomic chunks to maximize the context for the AI agent and for me personally to review every LOC}~(MS155).

Atomic did not necessarily mean small; participants saw it as context-dependent, provided the output stayed reviewable: \pq{as long as you review the full PR sometimes it's ok for it to be big, as long as testing is proportionally big} (MS19).

\textbf{(4)~When AI contributed, that its contributions be credited:}
\gl{CC}{6} was relevant to nearly all participants (R 95\%), but fewer wanted to adopt it in their own work (A 81\%), and participants mostly disagreed about what annotation would do.
Those who wanted to adopt it would \textit{mark AI's contributions where reviewers see them}, noting that \pq{knowing that code is AI-generated is critical to reviewing it more effectively} (MS185).

Those who were reluctant reasoned about ownership, wanting developers to stay accountable for their code: \pq{The more we annotate `this code was written by AI' the more we degrade the ownership\ldots we degrade accountability (not my fault, AI wrote that!)\ldots} (MS115).
Similarly, participants with more AI experience were less willing to adopt it (78\% vs.\ 86\%).
As MS19 put it: \pq{\ldots especially at this stage where it's assumed 90\% of what people do is AI generated\ldots I annotate my personal contributions instead}.
PF3 added that reviewers should instead \pq{judge all content on its merit}.


%% file: sec/5.Discussion.tex

\section{Discussion}
\label{sec:discussion}

Our aim was for the guidelines to capture what endures in developers' work as AI tools evolve.
They held up a year after we synthesized them, a considerable period given how rapidly AI tools are advancing, and with developers beyond the company and domain they came from, suggesting that they capture durable aspects of developers' work, although we do not claim that they apply universally (recall Sec.~\ref{sec:threats}).

\vspace{-1mm}
\subsection{\textbf{Guidelines as a shared vocabulary for engineering teams}}
\label{sec:disc-adapting}\label{sec:disc-vocabulary}

Participants' endorsements were not uniform, so no one set of guidelines is likely to fit every developer. We advocate accommodating this non-uniformity in developers' views on responsibly integrating AI.
Toward this end, our guidelines provide developers who hold differing views a shared vocabulary to discuss them, much as the cognitive dimensions framework gave designers shared terms for designing and discussing notations~\cite{green1998cognitive}.

For instance, one developer might hand AI their routine status updates, while a teammate might want to personally create their updates.
As we saw in \gl{MW}{2}, both can use the same terms: \enquote{routine communication and collaboration logistics}, which AI may automate, and \enquote{consequential stakeholder-facing messages}, which AI may not send without approval, with each individual deciding which ones are consequential.

We see this vocabulary as useful across three layers.
(1)~\textbf{Individuals} can compare the guidelines with their own practice to see which they already follow, which they would adopt if their work called for it, and which they resist, and why.
(2)~\textbf{Teams and organizations} can use them to set norms that fit their AI maturity and their domain; engineering leaders can bring them to a retrospective or a planning session as conversation starters, asking the same questions of the team, including where its members disagree.
The disagreements among individuals and within teams are often the most useful part, because they show where the work sets the boundary.
(3)~\textbf{Processes and tools} can build the guidelines into how work gets done, as agent skills do, while staying inclusive of developers across the spectrum.

These three layers can pull against each other where a developer's agency and their team's AI maturity diverge: a team where AI is not yet part of the work can hold back a developer ready to adopt a guideline, and a fast-moving team can set norms that some members are not ready, or do not want, to follow.
Because each guideline is deliberately general, teams should treat it as a default to confirm or revise for their context, leaving room for members who draw the line elsewhere.
Responsibly integrating AI, then, becomes a decision teams make together about where AI belongs in their work and where it does not.

\vspace{-2mm}
\subsection{\textbf{Guidelines as agent skills}}
\label{sec:disc-skills}

Developers increasingly steer AI agents through context files and skills, i.e., instructions checked into a repository that an agent loads when it starts or when a task calls for them~\cite{chatlatanagulchai2026agent,galster2026configuring}.
A skill encodes how a developer wants work done, and our guidelines have the potential to become such skills.


For example, \gl{DV}{4} could become a skill that loads whenever an agent changes code, splits the work into increments a developer can follow, and stops before applying a consequential change.
Similarly, \gl{QR}{4} could become a review skill that hands the agent's output to a different model, so that no agent certifies its own work (Sec.~\ref{subsec:res-qrm}).
Cross-cutting guidelines that apply to every task could be encoded as persistent agent instructions: e.g., asking before acting on missing context (\gl{CC}{1}), backing its claims with evidence (\gl{CC}{4}), and so on.
The demands would move into the agent, and developers could check that they were met.\looseness=-1

Because skills can also be files a team edits, the skills can carry the choices of Sec.~\ref{sec:disc-vocabulary}.
Within a team, one developer's setup might let the agent send routine build notifications, while another's routes every message to them as a draft (\gl{MW}{2}).
As another example, a team in a regulated domain might apply its sign-off skill (\gl{IO}{4}) to every production change, whereas another team might let agents apply routine configuration fixes but escalate the others.
Following research on scrutable personalization~\cite{kay2013scrutable}, these choices should remain transparent and steerable.

Overall, we see the success of the guidelines as agent skills as dependent on a shared vocabulary and on supporting developers’ diverse needs. Future research should examine whether this is the case and whether agents that follow these guidelines improve developers’ work outcomes and experiences.

%% file: sec/8.conclusion.tex
\section{Conclusion}
\label{sec:conclusion}

Software developers have been asking how to integrate AI into their daily work responsibly, and most have had to work out the answer on their own.
From the accounts of 860 developers who had begun working it out, we synthesized 35 developer-facing guidelines.
Participants described delegating to AI the work they could effectively verify and using it to grow their craft, while holding on to their judgment and the decisions they are accountable for.
For relationship-centric work, they drew the line differently, varying with their domain, SE, and AI experience.
A year later---a considerable time given AI tools' rapid advancement---we validated the guidelines derived from Phase~1 with an independent sample of developers, who strongly endorsed them, suggesting that these guidelines captured durable aspects of developers' work.

We invite teams to try these guidelines as a shared vocabulary: to choose those that fit their work and voice, settle openly where members differ, and externalize these choices so that peers and their AI agents can learn from them.

\vspace{1mm}

\noindent\textbf{Data Availability.} Our supplemental package is available at~\cite{supplemental}, and an interactive dashboard with an agent-skill generator is available at \href{https://aka.ms/ai-where-it-matters/guidelines}{\textcolor{blue}{\nolinkurl{aka.ms/ai-where-it-matters/guidelines}}}.

\noindent\textbf{Acknowledgments}. We thank Polis Georgiadis for distributing the survey at Pfizer and all participants for their insights.